\documentclass[10pt,conference]{IEEEtran}
\IEEEoverridecommandlockouts
\usepackage{soul,color}
\usepackage{cite}
\usepackage{amsmath,amssymb,amsfonts}
\usepackage{graphicx}
\usepackage{textcomp}
\usepackage{xcolor}
\usepackage{booktabs}
\usepackage{tikz}
\usetikzlibrary{arrows.meta}
\usepackage[hidelinks]{hyperref}
\usepackage{xspace}

\newif\ifcomment
\commenttrue
\ifcomment
  \newcommand{\sadra}[1]{{\color{red!70!black} [#1]$_\text{Sadra}$}}
  \newcommand{\rini}[1]{{\color{green!70!black} [#1]$_\text{Rini}$}}
  \newcommand{\hossein}[1]{{\color{blue!70!black} [#1]$_\text{Hossein}$}}
  \newcommand{\todo}[1]{{\color{orange!80!black} [TODO: #1]}}
\else
  \newcommand{\sadra}[1]{}
  \newcommand{\rini}[1]{}
  \newcommand{\hossein}[1]{}
  \newcommand{\todo}[1]{}
\fi

\newcommand{\nsp}{NSP\xspace}
\newcommand{\nsps}{NSPs\xspace}

\begin{document}

\title{FlashScript: Compiling Recurring Agent Workflows into Neuro-Symbolic Programs}

\author{
\IEEEauthorblockN{%
Sadra Sabouri\textsuperscript{1}\IEEEauthorrefmark{4},
Hosein Hasanbeig\textsuperscript{2},
Adam Sage\textsuperscript{2},
Shreya Sinha\textsuperscript{1},
Cale Darling\textsuperscript{2},\\
Ashish Tiwari\textsuperscript{2},
Katie Garms\textsuperscript{2},
Souti Chattopadhyay\textsuperscript{1},
Sumit Gulwani\textsuperscript{2}
}
\IEEEauthorblockA{\textsuperscript{1}University of Southern California, Los Angeles, CA\\
\{sabourih, ssinha05, schattop\}@usc.edu}
\IEEEauthorblockA{\textsuperscript{2}Microsoft, Redmond, WA\\
\{hhasanbeig, adamsage, cadarlin, astiwar, katwhite, sumitg\}@microsoft.com}
\thanks{\IEEEauthorrefmark{4}Part of this work was done during an internship at Microsoft.}
}

\maketitle

\begin{abstract}
Many agent tasks recur with new data or small changes, yet agents regenerate their reasoning and actions on each run, producing different results.
Distilling a prior run into a skill helps the agent avoid earlier mistakes, but it still re-runs the full process each time, so runs can vary and incur that cost.
We compile a successful agent run into a reusable neuro-symbolic program (NSP) with a set of guidelines we call \emph{FlashScript}.
Actions and data transformations become code, instance-specific values become parameters, and decisions that need model judgment become LLM calls.
We identify three layers where these calls belong and three techniques that replace most of them with code.
The compilation does not remove nondeterminism; it localizes it, so the deterministic program and its neural calls can be reviewed and tested separately.
Across 120 WorkBench analytics tasks with Claude Haiku~4.5, the \nsps were correct and made identical tool calls on every run, using about 70 times fewer input tokens and a third of the runtime of a skill-guided agent whose runs were inconsistent.
In a composite version of the tasks, the skill-guided agent's accuracy fell to 23\%, and its output inconsistency increased, while the \nsps remained correct and identical across all runs.
The \nsps successfully generalized to all 108 unseen instances of the same templates.
Our findings suggest that \nsps from agent trajectories can be considered cumulative software assets for end-users to test, maintain, and reuse in their recurring workflows.
\end{abstract}


\section{Introduction}

AI agents are increasingly performing recurring tasks on behalf of end users.
For example, an analyst asks an agent to update a weekly report using data from several sources without writing the code to retrieve, combine, and format it.
By translating natural-language requests into tool actions, agents advance the long-standing vision of end-user software engineering: letting people automate their own specialized needs~\cite{ko2011eusesurvey}.
Yet these agentic workflows do not behave like conventional software.
An agent reconstructs the procedure on every run, so repeated executions of the same request can diverge in path and outcome~\cite{yan2026inconlens,mehta2026agents}. 
A successful execution holds valuable procedural knowledge: which tools to use, what went wrong, and which decisions produced a desirable result~\cite{zhao2024expel,wang2025workflow}.
Currently, this knowledge is buried within the session: a user may inspect, correct, and approve a run, but that effort never carries over to future executions.
Users must therefore keep inspecting an already reviewed procedure.

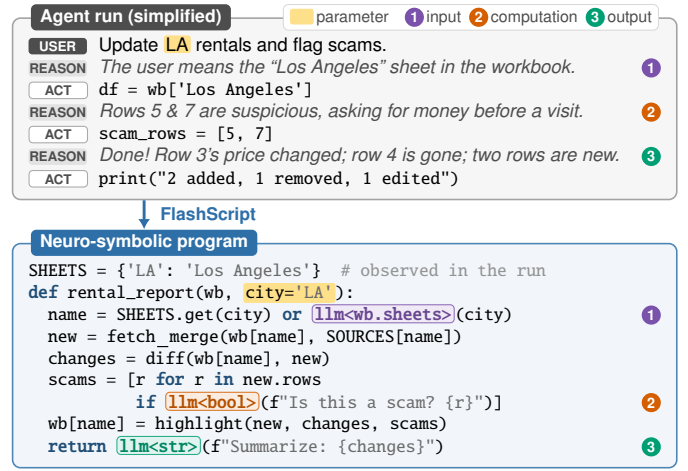
\begin{figure}[t]
\centering
\definecolor{tsAgent}{HTML}{8C8C8C}%
\definecolor{tsAgentDk}{HTML}{4D4D4D}%
\definecolor{tsProg}{HTML}{2E6FA8}%
\definecolor{tsIn}{HTML}{7B52AB}%
\definecolor{tsComp}{HTML}{D55E00}%
\definecolor{tsOut}{HTML}{009E73}%
\definecolor{tsParam}{HTML}{FFE08A}%
\newcommand{\tsTT}{\fontfamily{txtt}\selectfont}%
\newcommand{\tsLB}{\char`\{}%
\newcommand{\tsRB}{\char`\}}%
\newcommand{\tsUS}{\char`\_}%
\newcommand{\tsQ}{\textquotesingle}%
\newcommand{\tsI}{\phantom{xx}}%
\newcommand{\tsKw}[1]{{\bfseries\color{tsProg!70!black}#1}}%
\newcommand{\tsStr}[1]{{\color{black!58}#1}}%
\newcommand{\tsCmt}[1]{{\color{black!42}#1}}%
\newcommand{\tsThink}[1]{\leavevmode{\itshape\color{tsAgentDk}#1}}%
\DeclareRobustCommand{\tsMark}[2]{\tikz[baseline=(m.base)]\node[circle,fill=#2,text=white,%
  inner sep=0pt,minimum size=7.2pt,font=\sffamily\bfseries\fontsize{5.4}{5.4}\selectfont](m){#1};}%
\newcommand{\tsLLM}[2]{\tikz[baseline=(p.base)]\node[rounded corners=1.5pt,fill=#2!12,%
  draw=#2!85,line width=0.4pt,inner xsep=1pt,inner ysep=0.9pt,text=#2!85!black,%
  font=\tsTT\bfseries](p){llm<#1>};}%
\newcommand{\tsHL}[1]{\tikz[baseline=(h.base)]\node[fill=tsParam,rounded corners=1pt,%
  inner xsep=0.8pt,inner ysep=0.8pt](h){#1};}%
\newcommand{\tsTag}[2]{\tikz[baseline=(t.base)]\node[#2,rounded corners=1.4pt,inner xsep=0pt,%
  inner ysep=1.2pt,minimum width=22pt,font=\sffamily\bfseries\fontsize{5}{5}\selectfont](t){#1};}%
\newcommand{\tsUser}{\tsTag{USER}{fill=tsAgentDk,text=white}}%
\newcommand{\tsReason}{\tsTag{REASON}{fill=tsAgent!25,text=tsAgentDk}}%
\newcommand{\tsAct}{\tsTag{ACT}{draw=tsAgent!75,line width=0.4pt,fill=white,text=tsAgentDk}}%
\newsavebox{\tsRunBox}\newsavebox{\tsProgBox}\newsavebox{\tsBandBox}\newsavebox{\tsLegendBox}%
\sbox{\tsRunBox}{\begin{minipage}{\dimexpr\columnwidth-14pt\relax}%
\vspace*{3pt}\scriptsize\sffamily\renewcommand{\arraystretch}{1.04}%
\begin{tabular*}{\linewidth}{@{}l@{\hspace{4pt}}p{\dimexpr\linewidth-37pt\relax}@{\extracolsep{\fill}}r@{}}
\tsUser   & Update \tsHL{LA} rentals and flag scams. & \\
\tsReason & \tsThink{The user means the ``Los Angeles'' sheet in the workbook.} & \tsMark{1}{tsIn}\\
\tsAct    & \tsTT df = wb[\tsQ Los Angeles\tsQ] & \\
\tsReason & \tsThink{Rows 5 \& 7 are suspicious, asking for money before a visit.} & \tsMark{2}{tsComp}\\
\tsAct    & \tsTT scam\tsUS rows = [5, 7] & \\
\tsReason & \tsThink{Done! Row 3's price changed; row 4 is gone; two rows are new.} & \tsMark{3}{tsOut}\\
\tsAct    & \tsTT print("2 added, 1 removed, 1 edited") & \\
\end{tabular*}%
\end{minipage}}%
\sbox{\tsProgBox}{\begin{minipage}{\dimexpr\columnwidth-14pt\relax}%
\vspace*{3pt}\scriptsize\tsTT\renewcommand{\arraystretch}{1.02}%
\begin{tabular*}{\linewidth}{@{}l@{\extracolsep{\fill}}r@{}}
SHEETS = \tsLB\tsStr{\tsQ LA\tsQ}: \tsStr{\tsQ Los Angeles\tsQ}\tsRB\ \ \tsCmt{\# observed in the run} & \\
\tsKw{def} rental\tsUS report(wb, \tsHL{city=\tsStr{\tsQ LA\tsQ}}): & \\
\tsI name = SHEETS.get(city) \tsKw{or} \tsLLM{wb.sheets}{tsIn}(city) & \tsMark{1}{tsIn}\\
\tsI new = fetch\_merge(wb[name], SOURCES[name]) & \\
\tsI changes = diff(wb[name], new) & \\
\tsI scams = [r \tsKw{for} r \tsKw{in} new.rows & \\
\tsI\phantom{scams = [}\tsKw{if} \tsLLM{bool}{tsComp}(f\tsStr{"Is this a scam? \tsLB r\tsRB"})] & \tsMark{2}{tsComp}\\
\tsI wb[name] = highlight(new, changes, scams) & \\
\tsI \tsKw{return} \tsLLM{str}{tsOut}(f\tsStr{"Summarize: \tsLB changes\tsRB"}) & \tsMark{3}{tsOut}\\
\end{tabular*}%
\end{minipage}}%
\sbox{\tsBandBox}{\sffamily\bfseries\fontsize{6.6}{7.6}\selectfont\color{tsProg}FlashScript}%
\sbox{\tsLegendBox}{\sffamily\fontsize{6}{7}\selectfont\color{tsAgentDk}%
\tsHL{\phantom{LA}}\,parameter\enspace\enspace%
\tsMark{1}{tsIn}\,input\enspace\tsMark{2}{tsComp}\,computation\enspace\tsMark{3}{tsOut}\,output}%
\begin{tikzpicture}[
    panel/.style={line width=0.6pt,rounded corners=3pt,inner xsep=6pt,inner ysep=3.5pt},
    badge/.style={rounded corners=2pt,inner xsep=3.5pt,inner ysep=1.8pt,text=white,
                  font=\sffamily\bfseries\fontsize{6.6}{7}\selectfont}]
  \node[panel,draw=tsAgent!80,fill=tsAgent!8] (run) {\usebox{\tsRunBox}};
  \node[badge,fill=tsAgentDk,anchor=west] at ([xshift=7pt]run.north west) {Agent run (simplified)};
  \node[draw=black!15,line width=0.4pt,fill=white,rounded corners=2pt,inner xsep=2.5pt,
        inner ysep=1.2pt,anchor=east] at ([xshift=-7pt]run.north east) {\usebox{\tsLegendBox}};
  \node[panel,draw=tsProg!85,fill=tsProg!6,anchor=north] (prog) at ([yshift=-16pt]run.south)
       {\usebox{\tsProgBox}};
  \node[badge,fill=tsProg,anchor=west] (proglbl) at ([xshift=7pt]prog.north west)
       {Neuro-symbolic program};
  \draw[tsProg,line width=1.1pt,-{Latex[length=3.8pt,width=4.6pt]}]
       (proglbl.north |- run.south) -- (proglbl.north);
  \node[anchor=west,inner sep=0pt] at ([xshift=6pt,yshift=-5.5pt]proglbl.north |- run.south)
       {\usebox{\tsBandBox}};
\end{tikzpicture}
\caption{Compiling the agent's prior runs (top) into a neuro-symbolic program (bottom) using FlashScript.
The sheet selection process becomes a lookup with a neural fallback~\tsMark{1}{tsIn}, the scam judgment a yes-or-no call for each listing~\tsMark{2}{tsComp}, and the closing message a call that phrases changes computed by code~\tsMark{3}{tsOut}.
Colors mark the layer of each neural call.}
\label{fig:teaser}
\end{figure}

Prior work has attempted to preserve successful agent behavior in several forms.
Experience-based memory and workflow methods extract demonstrations, guidelines, or natural-language routines from past trajectories and retrieve them to guide later executions~\cite{zhao2024expel,fu2024autoguide,wang2025workflow}.
These methods help agents reuse experience, but the model must still interpret that guidance and decide how to execute it, so runs still vary.
At the other extreme, programming-by-demonstration and robotic process automation record and replay concrete action sequences~\cite{leshed2008coscripter,li2017sugilite,chasins2018rousillon,vanderaalst2018rpa}; they preserve execution more directly but generalize poorly when the inputs or environment change.
More recent methods compile agent traces into reusable artifacts: procedural skills in SKILL-DISCO, programmatic skills via neuro-symbolic learning, and mostly deterministic workflows in TraceCompiler~\cite{shao2026lifting,guo2026skill,yadouni2026tracecompiler,li2026preact,hasanbeig2026symbolic}.
These show the promise of trajectory compilation but leave open where the model's judgment must remain, and they do not address inconsistency.
What is missing is a consistent yet generalizable process that a user can rerun on each recurrence of their workflow.

In this paper, we explore how a recurring agent task can be preserved as a reusable \nsp (Fig.~\ref{fig:teaser}-bottom), with guidelines we call FlashScript.
A completed run is a natural starting point because users often reach it through several turns of feedback and manual fixes with an agent.
Our method builds on that interaction to create an \nsp.
The compilation turns actions into code, instance-specific values into parameters, and represents each decision that requires model judgment as a typed call \texttt{llm<T>}, which returns a value of declared type $T$, at a fixed point in the code.
We identify three layers where these calls belong, the input, computation, and output layers, and three techniques that replace most of them with code: caching observed values behind a guard, codifying judgments as rules, and clubbing independent questions into a call.
An \nsp would consistently follow the same steps, invoking the model only when needed to answer scoped questions.

We evaluated the resulting \nsps on 120 WorkBench analytics tasks~\cite{styles2024workbench} using Claude Haiku~4.5, against a vanilla agent and an agent with task-specific skills.
The \nsps were correct and made identical tool calls on every run, using about 70 times fewer input tokens and about a third of the runtime of either agent.
The skilled agent was correct in all three runs on 96\% of tasks, but on composite tasks, combining up to ten of these requests as subtasks, its accuracy fell to 23\%, while the \nsps stayed correct and identical in all runs.
The \nsps also solved all 108 unseen instances of the templates (100\% accuracy), against 96\% for the agent with skill.

\nsps can turn recurring agent work into specific software that end users can develop, bringing software engineering practices to programs including LLM calls~\cite{ko2011eusesurvey,mahmoudi2026smells}.
They do not eliminate nondeterminism, but they localize it.
Each neural call has an explicit input format, output type, and defined use in the \nsp.
This structure facilitates the separate testing and maintenance of the deterministic code and neural components.

\section{Motivating Example}
\label{sec:example}
Imagine an analyst who maintains a workbook of available rental listings for his clients each week.
Prices may change, some listings may be rented, and new rentals may become available.
Scam listings are also a big concern in large cities; therefore, he wants to help his clients avoid them.
So every Monday morning, he asks an agent to ``\emph{update LA rentals and flag scams}'' (Fig.~\ref{fig:teaser}-top).
The workbook has a separate sheet for each city.
The agent first infers that ``LA'' in the query refers to \texttt{Los Angeles} and selects the corresponding sheet.
It updates the workbook by fetching and merging new listings, then flags listings that request a money transfer before the first visit as potential scams.
Finally, it summarizes the changes.


What if he wants to repeat this next week or apply the same process for New York City?
He has already spent time clarifying the request and manually editing the workbook to get the output he wants for Los Angeles.
Ideally, the next run should be:
(C1) \emph{consistent}: the same output format, styling, and process as he reviewed and reached in the previous run;
(C2) \emph{generalizable}: it could handle variations in new data or in another city and reach the correct solution;
(C3) \emph{cheap}: reasonable costs for repeated runs, since the workflow has been refined and his corrections incorporated once, so the system should have already figured the solution out.


Our preliminary results show that a skill-guided agent does not fully meet these needs. 
In our sandbox, it generalizes to the New York City task, but across three runs it varies in how it names added columns and marks removed listings, taking an average of 38 seconds and using 42k tokens per run.
By contrast, the \nsp in Fig.~\ref{fig:teaser}-bottom flagged the same listings and produced identical sheets in 6.7 seconds using 6.7k tokens.

\section{FlashScript}
\label{sec:nsp}
FlashScript is our authoring guidance for writing these \nsps from prior agent runs.
Programs that combine code and invocations to learned models have a long history~\cite{chaudhuri2021neurosymbolic}, and LLMs have made them more common.
Models can serve as semantic predicates in synthesized programs~\cite{verbruggen2021semantic}, as functions in generated programs~\cite{gupta2023visprog,suris2023vipergpt,cheng2023binder}, and as operators in data pipelines~\cite{patel2024semanticoperators}.
We derive such an \nsp from a given user's session with an AI agent.
We turn a completed agent trajectory into a script with an additional operation, \texttt{llm<T>}, which sends a prompt to a model and returns a value of a declared type, such as a string, Boolean, or option.
The run’s actions become code, while its reasoning becomes model calls to be made when a future run requires judgment.

\emph{Fixing the Actions (C1).}
Each of the neural calls in \nsps is bounded.
The \nsp selects its inputs, checks that the response has the expected type~\cite{beurerkellner2023lmql,khattab2023dspy,willard2023outlines}, and makes the call at a fixed point in the code.
The \nsp's control flow determines when the model is queried, for what, and how its response is used.
As a result, variation across runs is confined to those calls.

\emph{Locating the Neural Part (C2).}
The calls fall into three layers, based on how information moves through the workflow.
\textbf{Input} layer calls turn unstructured text into typed values, such as identifying which sheet ``L.A.'' refers to.
\textbf{Computation} layer calls make semantic decisions that are difficult to express as rules, such as whether a listing is a scam.
\textbf{Output} layer calls render computed results in natural language for the task owner.

A program with no model calls is a fixed script: it is consistent, but limited only to the inputs it was generated from.
And a program that relies solely on model calls is equivalent to the reason-action agent loop.
\nsps sit between these two ends.
They keep actions as code and use the model calls only where expected.
Their generality depends on how many decisions they leave to model calls: delegating more handles more variation but reduces consistency and efficiency.

\emph{Efficient Decision Making (C3).}
Unlike an agent that makes an LLM call at every step, each reading the growing history, an \nsp pays only for its limited expected number of model calls.
Its expected cost depends on how often each call is made and how large it is.
We propose three techniques that can make calls less frequent or smaller, further reducing costs.
\textbf{Caching} keeps a value observed in the run and checks a condition before reusing it in a future run.
The model is called only if that condition fails.
For example, the sheet lookup in Fig.~\ref{fig:teaser} checks the \texttt{SHEETS} dictionary and falls back to an LLM call choosing from existing sheets.
\textbf{Codifying} replaces a repeatable judgment with a heuristic rule.
Writing the rule once costs less than asking the model every run and gives the same answer each time.
Unlike program-aided reasoning~\cite{gao2023pal,chen2022program}, the model does not make the decision by executing generated code.
In Fig.~\ref{fig:teaser}, a regex such as \texttt{(send|pay).*(Venmo|Zelle).*before.*} could flag the requests for suspicious payment before a visit, indicating potential scams.
\textbf{Clubbing} puts independent questions in one call or runs them in parallel~\cite{cheng2023batch,kim2024llmcompiler}.
These typed questions can also go to smaller classification models~\cite{chen2023frugalgpt,li2026jev}, since the generative capacity of an LLM is not required.

\section{Preliminary Results}
\label{sec:results}
We implemented \nsps as Python scripts with neural call using the same model as the agent.
We compared the \nsps with a vanilla agent and a skill-guided agent on the WorkBench benchmark~\cite{styles2024workbench} and evaluate whether they meet
consistency (C1),
generalization (C2),
and cost-effectiveness (C3)
criteria.

\emph{Setup.}
WorkBench's 690 tasks came from 69 request templates with ten instances each.
We took its 12 Analytics templates (120 tasks), in which the agent must read data, compute a value, compare it to a threshold, and decide whether and what to plot.
For example, the template ``\emph{If \{metric\} is more than \{threshold\} \{higher or lower\} than \{date\}, make a line plot of it since then}'' has the instance ``\emph{If engaged users is more than 10\% lower than November 24, make a line plot of it since then}.''
Other task categories are mostly single-record edits, leaving little room for runs to differ.
We first run the vanilla agent three times on every task.
Then for each template, WorkBench's grader selected the best run and a coding agent (Claude Code Opus 5.5) wrote the skill from that run.
The coding agent also wrote each \nsp from FlashScript instructions (Section~\ref{sec:nsp}) that provide the prompt template, the tool descriptions, and the same agent trajectory used to write the skill (the \emph{R\,+\,T} condition in Table~\ref{tab:ablation}).
We ran the skill-guided agent and the \nsp three times on every task.
All runs used Claude Haiku~4.5, chosen for its favorable performance–cost trade-off, at temperature~1, a common setting that reflects typical LLM generation behavior.

To examine how performance changes with task size, we combined requests from different templates into single prompts (``\emph{Please do all of the following: 1.\ \ldots\ 2.\ \ldots}''), using only instances not used to create the skill or \nsp.
We created ten composition tasks of varying lengths.
Their subtasks are independent, allowing the benchmark grader to evaluate each one separately.
Because errors can propagate when subtasks depend on one another, our setup likely understates the drop in accuracy on realistic tasks.

\begin{table}[t]
\centering
\caption{Performance metrics on repeated runs. Values are $\text{mean}\pm\text{std}$. the ``correct in all three runs'' row is the fraction of the tasks correct in every repeat.}
\label{tab:single}
\footnotesize
\setlength{\tabcolsep}{4pt}
\begin{tabular}{lrrr}
\toprule
 & Agent & Agent + skill & NSP \\
\midrule
Accuracy & 0.89$\pm$0.007 & 0.97$\pm$0.004 & \textbf{1.00$\pm$0} \\
Correct in all three runs & 0.79 & 0.96 & \textbf{1.00} \\
Same outcome as first run & 0.82$\pm$0.021 & 0.99$\pm$0.004 & \textbf{1.00$\pm$0} \\
Same tool calls as first run & 0.72$\pm$0.004 & 0.97$\pm$0.008 & \textbf{1.00$\pm$0} \\
Input tokens & 21,986$\pm$309 & 23,138$\pm$45 & \textbf{321$\pm$0} \\
Output tokens & 439$\pm$3.9 & 488$\pm$5.0 & \textbf{95$\pm$0.3} \\
Runtime (s) & 4.3$\pm$0.04 & 4.4$\pm$0.01 & \textbf{1.5$\pm$0.003} \\
\bottomrule
\end{tabular}
\end{table}

\begin{figure}[t]
\centering
\includegraphics[width=\columnwidth]{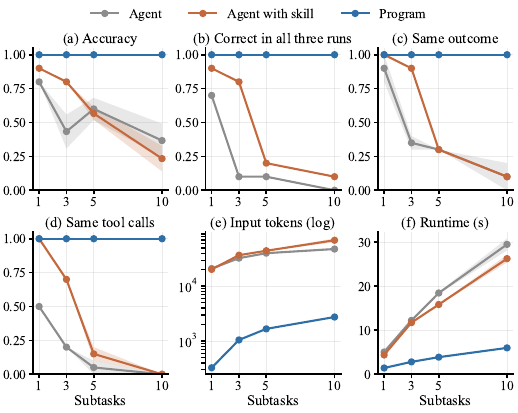}
\caption{Tasks made of 1, 3, 5, and 10 independent subtasks: ten tasks per length, three runs each. Bands show one standard deviation across the repeats.}
\label{fig:composite}
\end{figure}

\emph{Consistency (C1).}
The vanilla agent's runs were inconsistent: only 82\% of its repeated runs ended in the same state as the first run (Table~\ref{tab:single}).
The skill reduced most of this inconsistency but did not remove it completely (99\%).
The \nsp made the same tool calls in every run, and across runs only the wording of its reply varies.
On composite tasks, agents become less consistent (Fig.~\ref{fig:composite}b-f): with ten subtasks, the skilled agent reproduces the outcome of its first run in only 10\% of cases while working on the exact same task, receiving every subtask's skill and appearing to mix them up, adding plots no subtask requested and omitting some that were.
The \nsp, by contrast, produces the exact same result in every run.

\emph{Generalizability (C2).}
On repeated tasks, the vanilla agent was often correct (accuracy~0.89), and the skill reached 0.97.
They were correct in all three runs on 79\% and 96\% of tasks, respectively (Table~\ref{tab:single}).
Meanwhile, the \nsp was correct on every single run (accuracy~1.0).
Among the tasks, eight templates needed a decision about the data, such as whether a metric crossed a threshold or which date in the data ``Saturday'' refers to.
These templates account for 34 of the vanilla agent's 39 errors and all 11 of the skilled agent's.
The skilled agent fails in two ways.
When the skill was correct, the agent still made mistakes, for example by wrongly summing daily values in its reasoning ($2+3+6+3+8+5+\dots=148$ visits).
In other cases, the skill inherits a heuristic that was not generalizable beyond the reference run.
One template asks to compare a metric's percent growth, but the reference run compared their sums instead; both agreed there, so the grader accepted it, and the skill kept the shortcut, dropping accuracy to 0.67 against 0.87 for the vanilla agent on that task.

On composite tasks (ten per length, three runs each), the \nsp was correct in every run, compared with about 23\% for the skilled agent (Fig.~\ref{fig:composite}-a).
Across 108 unseen instances, the \nsp solved every task and repeated the first instance's tool sequence, against 96\% and 98\% for the skilled agent and 90\% and 78\% for the vanilla agent.
The \nsp also made no plot errors on composite tasks, while the skilled agent added 1.6 extra plots and omitted 0.6 plots per run compared with the anchor tasks, the same composites assembled from the first instance of each template.

\emph{Cost (C3).}
The \nsp used about 70 times fewer input tokens and about a third of the runtime of either agent (Table~\ref{tab:single}).
Across all \nsps there are 69 neural calls, 49 guarded input calls and 20 reply calls, averaging 1.5 per run.
On composite tasks its runtime grows from 1.4 to 6.0 seconds, against about 26 to 30 for the agents.
All three reduction techniques from Section~\ref{sec:nsp} helped: caching served 37 of the 49 guarded input calls from a stored value instead of the model, codifying turned every computation-layer decision into a rule, and clubbing combined independent questions into one call, cutting a ten-subtask task from 5.8 model questions to 2.0.

We also studied how an NSP's quality depends on the generation instructions and on the trajectory.
Our first instructions asked the agent only to write an \nsp from a template or a trajectory; those \nsps achieved 0.96 and 0.88 accuracy, respectively, but failed on reworded requests and edge cases.
Accuracy reached 1.00 once we asked the agent to plan first, encode decisions as rules, and place calls according to their layer (available in supplementary materials\footnote{https://figshare.com/s/c812d719d2d43bae034c}).
With these instructions, we wrote \nsps from three inputs: the template only (T), the run alone (R), and the template with the skill's reference run (R+T) (Table~\ref{tab:ablation}).
Only using the run makes the \nsp too specific.
It is correct on the instance it is generated from, but fails on unseen instances.
Using the template alone already yields good generalization.
It identifies the variables and plans for the variants.
Adding the run to the template strengthens the NSP's caches by setting more inclusive guards, reducing the number of calls from 1.8 to 1.5.
The run also exposes the agent's decisions to review.
For example, the plan caught the reference run's shortcut for summing and corrected it.
In short, correctness comes from the template and the planning instructions, while the run contributes efficiency through tighter caches and auditability by exposing the reference run's decisions for review.

\begin{table}[t]
\centering
\caption{Performance metrics for \nsps written from the reference run (R), the request template (T), and both (R+T). Composite values are averaged over the four lengths.}
\label{tab:ablation}
\footnotesize
\setlength{\tabcolsep}{4pt}
\begin{tabular}{lrrr}
\toprule
 & R & T & R+T \\
\midrule
Accuracy, same instance & 1.00 & 1.00 & 1.00 \\
Accuracy, unseen instances & 0.89 & 1.00 & 1.00 \\
Accuracy, composite tasks & 0.53 & 1.00 & 1.00 \\
Same outcome, composite tasks & 0.99 & 1.00 & 1.00 \\
Runs that raise an error & 8.6\% & 0\% & 0\% \\
Neural calls per run & 1.3 & 1.8 & 1.5 \\
\bottomrule
\end{tabular}
\end{table}


\section{Discussion}
\nsps also changes what users retain after a successful run: reusable software that can be parameterized, combined with other \nsps, and maintained over time.

\emph{User Interaction with NSPs.}
Instead of having an agent solve the same task from scratch each time, the user keeps an \nsp they reuse.
An \nsp can expose user-facing parameters for the task.
For a scam check, the user could choose a heuristic from a menu or describe what they consider a scam as a text input.
The \nsp includes that preference in its LLM scam-check call, grounding the model’s judgment in the user’s criteria.
The user can change the preference and see its effect on the output within seconds.

\emph{Software Lifecycle and Testing.}
Because the compiled control flow is fixed and separates deterministic code from model calls, making LLM-based systems easier to analyze and validate~\cite{scotti2026tamed}, \nsps can follow the same lifecycle as other software.
Developers can inspect the control flow before deployment, see which steps run and where the model is called, and remove the ``black-box'' risk of an agent.
Testing then splits by component: the symbolic shell takes the usual unit, integration, and regression tests; the neural calls, whose outputs vary across runs and models, need contract, repeated-sampling, and metamorphic tests; and the seam must check the shell handles any value of a call's return type $T$.
As a workflow settles, refactoring, such as rule codification, replaces neural calls with deterministic code.

\emph{Benchmarking Nuanced Decisions.}
Agent benchmarks favor tasks with clear, verifiable outcomes, so they rarely capture the nuanced decisions involved in recurring work.
Real tasks may require models to infer constraints, interpret vague terms such as ``suspicious,'' weigh personal preferences, or frame open-ended problems~\cite{simon1947administrative,simon1973illstructured,keeney1976decisions,simon1956rational,hullermeier2024preference,bankes1993exploratory,lempert2003shaping,checkland1981systems,baradaran2026llm}. A dataset of such workflows would clarify the boundary of when an \nsp neural calls.



\section{Future Plans}
\label{sec:future}
We will build a benchmark of recurring workflows that test the nuanced decisions an agent must make before acting. Most benchmarks focus on whether the agent uses the right tools, even if it misunderstood the task. Our benchmark will test whether it can infer unstated constraints, define vague terms such as ``suspicious''~\cite{baradaran2026llm}, weigh personal preferences, and decide what to investigate when there is no clear path or success measure.
Each workflow will include successive instances, with real users’ judgments as ground truth.
We will test how well caches and guards handle changing data, and measure each model call’s accuracy and consistency in a \nsp separately from task execution.

Additionally, we will study interfaces that let end-user developers inspect and customize \nsps without reading code.
Users could set text or menu parameters, edit steps or model guidance, and see the effect on outputs instantly.
Further user studies will explore which parameters to expose, how users want to set them for an \nsp, and what they expect \nsps to generalize across recurring runs.

Finally, we will make \nsp coverage visible for the user. For each new request, model calls will identify its requirements and check which ones the \nsp can handle. Covered requirements will run as code, and a fallback agentic step will handle the rest and can update the \nsp.
Over time, the \nsp can cover more scenarios and data structures, so more of each request can run as reviewed code.

\bibliographystyle{IEEEtran}
\bibliography{references}

@article{beurerkellner2023lmql,
  title={Prompting Is Programming: A Query Language for Large Language Models},
  author={Beurer-Kellner, Luca and Fischer, Marc and Vechev, Martin},
  journal={Proceedings of the ACM on Programming Languages},
  volume={7},
  number={PLDI},
  year={2023},
  doi = {10.1145/3591300}
}

@inproceedings{chasins2018rousillon,
  title={Rousillon: Scraping Distributed Hierarchical Web Data},
  author={Chasins, Sarah E. and Mueller, Maria and Bodik, Rastislav},
  booktitle={Proceedings of the ACM Symposium on User Interface Software and Technology (UIST)},
  year={2018},
  doi = {10.1145/3242587.3242661}
}

@article{chaudhuri2021neurosymbolic,
  title={Neurosymbolic Programming},
  author={Chaudhuri, Swarat and Ellis, Kevin and Polozov, Oleksandr and Singh, Rishabh and Solar-Lezama, Armando and Yue, Yisong},
  journal={Foundations and Trends in Programming Languages},
  volume={7},
  number={3},
  year={2021},
  doi = {10.1561/2500000049}
}

@article{chen2022program,
  title={Program of thoughts prompting: Disentangling computation from reasoning for numerical reasoning tasks},
  author={Chen, Wenhu and Ma, Xueguang and Wang, Xinyi and Cohen, William W},
  journal={Transactions on Machine Learning Research},
  year={2023},
  doi = {10.48550/arXiv.2211.12588}
}

@article{chen2023frugalgpt,
  title={{FrugalGPT}: How to Use Large Language Models While Reducing Cost and Improving Performance},
  author={Chen, Lingjiao and Zaharia, Matei and Zou, James},
  journal={Transactions on Machine Learning Research},
  year={2024},
  doi = {10.48550/arXiv.2305.05176}
}

@inproceedings{cheng2023batch,
  title={Batch Prompting: Efficient Inference with Large Language Model {APIs}},
  author={Cheng, Zhoujun and Kasai, Jungo and Yu, Tao},
  booktitle={Proceedings of the 2023 Conference on Empirical Methods in Natural Language Processing: Industry Track},
  year={2023},
  doi = {10.18653/v1/2023.emnlp-industry.74}
}

@inproceedings{cheng2023binder,
  title={Binding Language Models in Symbolic Languages},
  author={Cheng, Zhoujun and Xie, Tianbao and Shi, Peng and Li, Chengzu and Nadkarni, Rahul and Hu, Yushi and Xiong, Caiming and Radev, Dragomir and Ostendorf, Mari and Zettlemoyer, Luke and Smith, Noah A. and Yu, Tao},
  booktitle={International Conference on Learning Representations},
  year={2023},
  doi = {10.48550/arXiv.2210.02875}
}

@inproceedings{gao2023pal,
  title={{PAL}: Program-aided Language Models},
  author={Gao, Luyu and Madaan, Aman and Zhou, Shuyan and Alon, Uri and Liu, Pengfei and Yang, Yiming and Callan, Jamie and Neubig, Graham},
  booktitle={International Conference on Machine Learning},
  pages={10764--10799},
  year={2023},
  doi = {10.48550/arXiv.2211.10435}
}

@inproceedings{gupta2023visprog,
  title={Visual Programming: Compositional Visual Reasoning Without Training},
  author={Gupta, Tanmay and Kembhavi, Aniruddha},
  booktitle={Proceedings of the IEEE/CVF Conference on Computer Vision and Pattern Recognition (CVPR)},
  year={2023},
  doi = {10.1109/CVPR52729.2023.01436}
}

@inproceedings{khattab2023dspy,
  title={{DSPy}: Compiling Declarative Language Model Calls into State-of-the-Art Pipelines},
  author={Khattab, Omar and Singhvi, Arnav and Maheshwari, Paridhi and Zhang, Zhiyuan and Santhanam, Keshav and Vardhamanan, Sri and Haq, Saiful and Sharma, Ashutosh and Joshi, Thomas T. and Moazam, Hanna and Miller, Heather and Zaharia, Matei and Potts, Christopher},
  journal={The Twelfth International Conference on Learning Representations},
  year={2024},
  doi = {10.48550/arXiv.2310.03714}
}

@inproceedings{kim2024llmcompiler,
  title={An {LLM} Compiler for Parallel Function Calling},
  author={Kim, Sehoon and Moon, Suhong and Tabrizi, Ryan and Lee, Nicholas and Mahoney, Michael W. and Keutzer, Kurt and Gholami, Amir},
  booktitle={International Conference on Machine Learning},
  year={2024},
  doi = {10.48550/arXiv.2312.04511}
}

@inproceedings{leshed2008coscripter,
  title={{CoScripter}: Automating \& Sharing How-To Knowledge in the Enterprise},
  author={Leshed, Gilly and Haber, Eben M. and Matthews, Tara and Lau, Tessa},
  booktitle={Proceedings of the SIGCHI Conference on Human Factors in Computing Systems (CHI)},
  year={2008},
  doi = {10.1145/1357054.1357323}
}

@inproceedings{li2017sugilite,
  title={{SUGILITE}: Creating Multimodal Smartphone Automation by Demonstration},
  author={Li, Toby Jia-Jun and Azaria, Amos and Myers, Brad A.},
  booktitle={Proceedings of the CHI Conference on Human Factors in Computing Systems},
  year={2017},
  doi = {10.1145/3025453.3025483}
}

@article{li2026jev,
  title={{JEV}-as-a-Judge: Accept When Confident, Escalate When Unsure},
  author={Li, Yubo and Miao, Yidi and Krishnan, Ramayya and Padman, Rema},
  journal={arXiv preprint arXiv:2609.26550},
  year={2026},
  doi = {10.48550/arXiv.2609.26550}
}

@article{mehta2026agents,
  title={When Agents Disagree With Themselves: Behavioral Consistency as an Uncertainty Signal for {LLM} Agents},
  author={Mehta, Aman},
  journal={arXiv preprint arXiv:2602.11619},
  year={2026},
  doi = {10.48550/arXiv.2602.11619}
}

@article{patel2024semanticoperators,
  title={Semantic Operators: A Declarative Model for Rich, {AI}-based Data Processing},
  author={Patel, Liana and Jha, Siddharth and Pan, Melissa and Gupta, Harshit and Asawa, Parth and Guestrin, Carlos and Zaharia, Matei},
  journal={arXiv preprint arXiv:2407.11418},
  year={2024},
  doi = {10.48550/arXiv.2407.11418}
}

@article{styles2024workbench,
  title={{WorkBench}: a Benchmark Dataset for Agents in a Realistic Workplace Setting},
  author={Styles, Olly and Miller, Sam and Cerda-Mardini, Patricio and Guha, Tanaya and Sanchez, Victor and Vidgen, Bertie},
  journal={arXiv preprint arXiv:2405.00823},
  year={2024},
  doi = {10.48550/arXiv.2405.00823}
}

@inproceedings{suris2023vipergpt,
  title={{ViperGPT}: Visual Inference via Python Execution for Reasoning},
  author={Sur{\'\i}s, D{\'\i}dac and Menon, Sachit and Vondrick, Carl},
  booktitle={Proceedings of the IEEE/CVF International Conference on Computer Vision (ICCV)},
  year={2023},
  doi = {10.1109/ICCV51070.2023.01092}
}

@article{vanderaalst2018rpa,
  title={Robotic Process Automation},
  author={van der Aalst, Wil M. P. and Bichler, Martin and Heinzl, Armin},
  journal={Business \& Information Systems Engineering},
  volume={60},
  number={4},
  year={2018},
  doi = {10.1007/s12599-018-0542-4}
}

@article{verbruggen2021semantic,
  title={Semantic programming by example with pre-trained models},
  author={Verbruggen, Gust and Le, Vu and Gulwani, Sumit},
  journal={Proceedings of the ACM on Programming Languages},
  volume={5},
  number={OOPSLA},
  pages={1--25},
  year={2021},
  publisher={ACM New York, NY, USA},
  doi = {10.1145/3485477}
}

@article{willard2023outlines,
  title={Efficient Guided Generation for Large Language Models},
  author={Willard, Brandon T. and Louf, R{\'e}mi},
  journal={arXiv preprint arXiv:2307.09702},
  year={2023},
  doi = {10.48550/arXiv.2307.09702}
}

@article{yan2026inconlens,
  title={{InconLens}: Interactive Visual Diagnosis of Behavioral Inconsistencies in {LLM}-based Agentic Systems},
  author={Yan, Shuo and Wen, Xiaolin and Ruan, Shaolun and Zhang, Yanjie and Mi, Jiaming and Sun, Yushi and Qu, Huamin and Sheng, Rui},
  journal={arXiv preprint arXiv:2603.28106},
  year={2026},
  doi = {10.48550/arXiv.2603.28106}
}

@inproceedings{zhao2024expel,
  title={{ExpeL}: {LLM} Agents Are Experiential Learners},
  author={Zhao, Andrew and Huang, Daniel and Xu, Quentin and Lin, Matthieu and Liu, Yong-Jin and Huang, Gao},
  booktitle={Proceedings of the AAAI Conference on Artificial Intelligence},
  year={2024},
  doi = {10.1609/aaai.v38i17.29936}
}

@inproceedings{mahmoudi2026smells,
  title={Specification and Detection of {LLM} Code Smells},
  author={Mahmoudi, Brahim and Chenail-Larcher, Zacharie and Moha, Naouel and Sti{\'e}venart, Quentin and Avellaneda, Florent},
  booktitle={Proceedings of the 48th International Conference on Software Engineering: New Ideas and Emerging Results (ICSE-NIER)},
  pages={181--185},
  year={2026},
  doi = {10.1145/3786582.3786835}
}

@inproceedings{baradaran2026llm,
  title={{LLM}-Integrated Declarative Program Analysis},
  author={Baradaran, Sara and Nazari, Amirmohammad and Raghothaman, Mukund},
  booktitle={Proceedings of the 15th ACM SIGPLAN International Workshop on the State Of the Art in Program Analysis},
  pages={41--47},
  year={2026},
  doi = {10.1145/3814987.3814993}
}

@article{ko2011eusesurvey,
  author  = {Ko, Amy J. and Abraham, Robin and Beckwith, Laura and
             Blackwell, Alan and Burnett, Margaret and Erwig, Martin and
             Scaffidi, Chris and Lawrance, Joseph and Lieberman, Henry and
             Myers, Brad and Rosson, Mary Beth and Rothermel, Gregg and
             Shaw, Mary and Wiedenbeck, Susan},
  title   = {The State of the Art in End-User Software Engineering},
  journal = {ACM Computing Surveys},
  volume  = {43},
  number  = {3},
  articleno = {21},
  pages   = {1--44},
  year    = {2011},
  month   = apr,
  doi     = {10.1145/1922649.1922658}
}

@inproceedings{wang2025workflow,
  author    = {Wang, Zora Zhiruo and Mao, Jiayuan and
               Fried, Daniel and Neubig, Graham},
  title     = {Agent Workflow Memory},
  booktitle = {Proceedings of the 42nd International Conference on Machine Learning},
  series    = {Proceedings of Machine Learning Research},
  volume    = {267},
  pages     = {63897--63911},
  publisher = {PMLR},
  year      = {2025},
  url       = {https://proceedings.mlr.press/v267/wang25bx.html},
  doi = {10.48550/arXiv.2409.07429}
}

@article{scotti2026tamed,
  author    = {Scotti, Vincenzo and Keim, Jan and Hey, Tobias and
               Metzger, Andreas and Koziolek, Anne and
               Mirandola, Raffaela},
  title     = {A Road Map for Tamed Interactions with Large Language Models},
  journal   = {ACM Transactions on Software Engineering and Methodology},
  volume    = {35},
  number    = {9},
  articleno = {279},
  pages     = {1--40},
  year      = {2026},
  month     = aug,
  publisher = {Association for Computing Machinery},
  doi       = {10.1145/3832776},
  url       = {https://doi.org/10.1145/3832776}
}

@article{fu2024autoguide,
  title={Autoguide: Automated generation and selection of context-aware guidelines for large language model agents},
  author={Fu, Yao and Kim, Dong-Ki and Kim, Jaekyeom and Sohn, Sungryull and Logeswaran, Lajanugen and Bae, Kyunghoon and Lee, Honglak},
  journal={Advances in Neural Information Processing Systems},
  volume={37},
  pages={119919--119948},
  year={2024},
  doi = {10.48550/arXiv.2403.08978}
}

@article{hasanbeig2026symbolic,
  title={Symbolic task inference in deep reinforcement learning},
  author={Hasanbeig, Hosein and Jeppu, Natasha Yogananda and Abate, Alessandro and Melham, Tom and Kroening, Daniel},
  journal={Journal of Artificial Intelligence Research},
  volume={80},
  pages={1099--1137},
  year={2024},
  doi={10.1613/jair.1.14063}
}

@article{shao2026lifting,
  title={Lifting traces to logic: Programmatic skill induction with neuro-symbolic learning for long-horizon agentic tasks},
  author={Shao, Jie-Jing and Yin, Haiyan and Lyu, Yueming and Yu, Xingrui and Guo, Lan-Zhe and Tsang, Ivor and Kwok, James and Li, Yu-Feng},
  journal={arXiv preprint arXiv:2605.01293},
  year={2026},
  doi = {10.48550/arXiv.2605.01293}
}

@article{guo2026skill,
  title={SKILL-DISCO: Distilling and Compiling Agent Traces into Reusable Procedural Skills},
  author={Guo, Zhongxin and Qi, Danrui and Gu, Hanwen and Cheng, Peng and Xiong, Yongqiang},
  journal={arXiv preprint arXiv:2606.26669},
  year={2026},
  doi = {10.48550/arXiv.2606.26669}
}

@article{li2026preact,
  title={PreAct: Computer-Using Agents that Get Faster on Repeated Tasks},
  author={Li, Bojie},
  journal={arXiv preprint arXiv:2606.17929},
  year={2026},
  doi = {10.48550/arXiv.2606.17929}
}

@article{yadouni2026tracecompiler,
  title={TraceCompiler: Skill-Guided Mining and Compilation of LLM Agent Traces into Mostly Deterministic Workflows},
  author={Yadouni, Salma El and Li, Guanyi},
  journal={arXiv preprint arXiv:2608.02680},
  year={2026},
  doi = {10.48550/arXiv.2608.02680}
}

@book{simon1947administrative,
  title={Administrative Behavior: A Study of Decision-Making Processes in Administrative Organization},
  author={Simon, Herbert A.}, year={1947}, publisher={Macmillan}}

@article{simon1956rational,
  title={Rational choice and the structure of the environment},
  author={Simon, Herbert A.}, journal={Psychological Review},
  volume={63}, number={2}, pages={129--138}, year={1956},
  doi = {10.1037/h0042769}
}

@article{simon1973illstructured,
  title={The structure of ill-structured problems},
  author={Simon, Herbert A.}, journal={Artificial Intelligence},
  volume={4}, number={3--4}, pages={181--201}, year={1973},
  doi = {10.1016/0004-3702(73)90011-8}
}

@book{keeney1976decisions,
  title={Decisions with Multiple Objectives: Preferences and Value Trade-offs},
  author={Keeney, Ralph L. and Raiffa, Howard}, year={1976}, publisher={John Wiley \& Sons},
  doi = {10.1017/CBO9781139174084}
}

@article{hullermeier2024preference,
  title={Preference learning and multiple criteria decision aiding: Differences, commonalities, and synergies---Part {I}},
  author={H{\"u}llermeier, Eyke and S{\l}owi{\'n}ski, Roman},
  journal={4OR}, volume={22}, pages={179--209}, year={2024},
  number = {2},
  doi = {10.1007/s10288-023-00560-6}
}

@article{bankes1993exploratory,
  title={Exploratory modeling for policy analysis},
  author={Bankes, Steve}, journal={Operations Research},
  volume={41}, number={3}, pages={435--449}, year={1993},
  doi = {10.1287/opre.41.3.435}
}

@book{lempert2003shaping,
  title={Shaping the Next One Hundred Years: New Methods for Quantitative, Long-Term Policy Analysis},
  author={Lempert, Robert J. and Popper, Steven W. and Bankes, Steven C.},
  year={2003}, publisher={RAND Corporation},
  doi = {10.7249/MR1626}
}

@book{checkland1981systems,
  title={Systems Thinking, Systems Practice},
  author={Checkland, Peter}, year={1981}, publisher={John Wiley \& Sons}}

\end{document}